\documentclass[conference]{IEEEtran}

\IEEEoverridecommandlockouts
\usepackage{epsfig}
\usepackage{graphicx}
\usepackage{amsmath,amssymb,amsfonts}
\usepackage{algorithmic}
\usepackage{graphicx}
\usepackage{textcomp}
\usepackage{xcolor}
\usepackage{url}
\usepackage{booktabs}
\usepackage{flushend}
\def\BibTeX{{\rm B\kern-.05em{\sc i\kern-.025em b}\kern-.08em
    T\kern-.1667em\lower.7ex\hbox{E}\kern-.125emX}}

\usepackage{fancyhdr}

\usepackage{mathtools,subcaption}

\DeclareMathOperator*{\argmin}{arg\,min}

\fancypagestyle{copyright}{
    \fancyhf{}
    \fancyhead[C]{
        \small
        \textit{IEEE International Conference on Multimedia Information Processing and Retrieval (MIPR)}, San Jose, CA, 2024. \\
        \vspace{5pt} \scriptsize
        © 2024 IEEE. Personal use of this material is permitted. Permission from IEEE must be obtained for all other uses, in any current or future media, including reprinting/republishing this material for advertising or promotional purposes, creating new collective works, for resale or redistribution to servers or lists, or reuse of any copyrighted component of this work in other works.
    }
    
}

\begin{document}

\title{\vspace{5pt}Mutual Information Analysis in Multimodal Learning Systems}

\author{Hadi Hadizadeh,$^{1}$ S. Faegheh Yeganli,$^{1}$ Bahador Rashidi,$^{2}$ and Ivan V. Baji\'{c}$^{1}$\thanks{This work was supported by Huawei Canada, Digital Research Alliance of Canada, and NSERC grants RGPIN-2021-02485 and RGPAS-2021-00038.}\\
$^{1}$Simon Fraser University, Burnaby, BC, Canada \hspace{20pt}$^{2}$Huawei Canada, Edmonton, AB, Canada\\
{\tt\small hadi\_hadizadeh@sfu.ca, syeganli@sfu.ca, bahador.rashidi@huawei.com, ibajic@ensc.sfu.ca}
\vspace{-5pt}
}

\maketitle
\thispagestyle{copyright}

\begin{abstract}
In recent years, there has been a significant increase in applications of multimodal signal processing and analysis, largely driven by the increased availability of multimodal datasets and the rapid progress in multimodal learning systems. Well-known examples include autonomous vehicles, audiovisual generative systems, vision-language systems, and so on. Such systems integrate multiple signal modalities: text, speech, images, video, LiDAR, etc., to perform various tasks. A key issue for understanding such systems is the relationship between various modalities and how it impacts task performance. In this paper, we employ the concept of mutual information (MI) to gain insight into this issue. Taking advantage of the recent progress in entropy modeling and estimation, we develop a system called InfoMeter to estimate MI between modalities in a multimodal learning system. We then apply InfoMeter to analyze a multimodal 3D object detection system over a large-scale dataset for autonomous driving. Our experiments on this system suggest that a lower MI between modalities is beneficial for detection accuracy. This new insight may facilitate improvements in the development of future multimodal learning systems. 
  
\end{abstract}

\vspace{5pt}


\section{Introduction}
\label{sec:intro}
With the rapid advancement of artificial intelligence (AI) technologies and the growth of data generated across various platforms and domains, ranging from social media to healthcare, multimodal learning has emerged as a powerful and popular paradigm for data processing and analysis in diverse applications such as autonomous robots and vehicles~\cite{bojarski2016end}, large language models~\cite{bubeck2023sparks}, and so on. 
In such applications, data from multiple modalities (e.g., text, image, audio, video) are integrated to perform certain tasks. Multimodal systems offer the possibility of expanding the performance beyond that of unimodal systems by leveraging information available in multiple modalities. A key issue in understanding the operation of multimodal systems is the relationship between modalities and how such a relationship affects performance. 

A widespread belief in the multimodal machine learning research community is that modalities should ``reinforce'' each other~\cite{lv2021cvpr,xu2022access}. Recently, transformers have been a popular approach for this purpose~\cite{xu2023tpami}. From this point of view, it would seem beneficial if each modality carried a lot of information about the other(s). On the other hand, one could also argue that a modality that carries a lot of information about another modality is redundant, and it is not clear how this would be beneficial to the multimodal learning system. One way to quantify redundancy is through the concept of mutual information (MI)~\cite{Cover}, which represents the (Shannon) information that one random variable carries about another. The ``reinforcement'' argument noted above suggests that high MI between modalities is beneficial in multimodal learning systems, whereas the ``redundancy'' argument might suggest otherwise. In this paper, we examine this issue by focusing on the 3D object detection from camera and LiDAR modalities, which is a representative multimodal computer vision problem in autonomous driving. 

MI estimation in high-dimensional spaces is very challenging due to the ``curse of dimensionality''~\cite{bellman1957dynamic}, which is the need for exponentially more data as the dimension increases in order to provide reliable probability estimates. Nonetheless, due to the importance of estimating MI empirically from data, many MI estimators have been proposed in the literature~\cite{parzen1962estimation,silverman1986density,liam,mcallester20aistats,belghazi2021mine,kraskov2004estimating}, relying on concepts such as kernel density estimation, variational approaches, or estimating bounds on MI. The approach we take in this paper is to leverage state-of-the-art entropy estimators developed in the area of learning-based compression~\cite{balle2018, minnen, cheng2020}, and use the relationship between entropy and MI to provide an estimate. In this way, we develop a system called InfoMeter (``Information Meter''), which can be used to estimate MI between any two points in an information processing pipeline, much like a voltmeter can be used to measure voltage between any two points in an electical circuit. Using the InfoMeter, we then analyze the performance of a multimodal 3D object detection system as a function of the MI between the camera and LiDAR modalities. To our knowledge, such analysis has not been performed before. The analysis reveals that the lower the MI between the modalities during training, the better the accuracy of the final trained system. This lends support to the ``redundancy'' argument mentioned above. 

The paper is organized as follows. In Section~\ref{sec:MI_multimodal}, we present the InfoMeter and explain how it was used to estimate MI in a multimodal system for 3D object recognition. Experiments are described and analyzed in Sections~\ref{sec:experiments} and~\ref{sec:analysis}, respectively, followed by conclusions in Section~\ref{sec:conclusions}.

\begin{figure}
\centering
\includegraphics[width=\columnwidth]{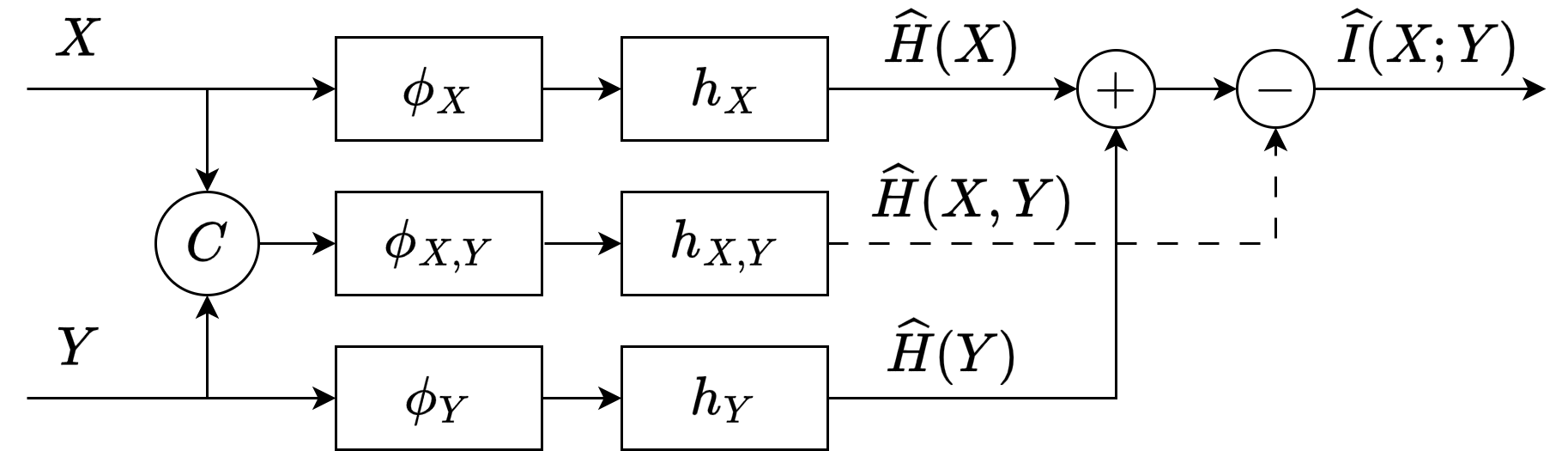}
\caption{Block diagram of the InfoMeter for estimating the MI between sources $X$ and $Y$. $C$ represents concatenation, $\phi$'s are invertible transformations and $h$'s are entropy estimators. }
\label{fig:infometer}
\end{figure}

\section{Mutual Information in Multimodal Systems}
\label{sec:MI_multimodal}
In this section, we first present the InfoMeter -- the proposed system for estimatting the mutual information (MI) -- and then describe how InfoMeter was used for MI analysis in a multimodal system for 3D object detection.  

\subsection{InfoMeter}
\label{sec:InfoMeter}

The block diagram of the InfoMeter is shown in Fig.~\ref{fig:infometer}, where the goal is to estimate the MI between sources $X$ and $Y$. The idea is to use the basic relationship between the MI and entropy~\cite{Cover},
\begin{equation}
\label{eq:MI}
I(X;Y) = H(X) + H(Y) - H(X,Y),
\end{equation}
and leverage recent advances in entropy estimation from learning-based compression to estimate the three entropy terms on the right-hand side. Specifically, $X$ and $Y$ are first passed through learnable invertible transformations $\phi_X$ and $\phi_Y$, respectively, and then fed to their respective entropy estimators $h_X$ and $h_Y$. The reason is that each entropy estimator assumes a certain probabilistic model for its input, for example, factorized~\cite{factorized}, hyperprior~\cite{balle2018}, autoregressive~\cite{minnen}, etc., and it is important to transform the actual input into the latent probabilistic form assumed by the estimator. At the same time, the transformation should be invertible, since such transformations preserve the entropy:
\begin{equation}
    H(\phi_X(X))=H(X), \qquad H(\phi_Y(Y))=H(Y).
\end{equation}

In parallel, $X$ and $Y$ are concatenated into $(X,Y)$, and this concatenation is passed through its own learnable invertible transformation $\phi_{X,Y}$ and its entropy estimator $h_{X,Y}$. In the end, three entropy estimates are produced: $\widehat{H}(X)$, $\widehat{H}(Y)$, and $\widehat{H}(X,Y)$, and these are combined using~(\ref{eq:MI}) to arrive at the final MI estimate:
\begin{equation}
    \widehat{I}(X;Y) = \widehat{H}(X) + \widehat{H}(Y) - \widehat{H}(X,Y).
    \label{eq:MI_estimate}
\end{equation}

The approach described above is fairly generic. $X$ and $Y$ could be any sources of information -- for example, two points in a neural network -- and InfoMeter could be used as a probe to estimate the MI between them. In the remainder of the paper, we apply InfoMeter to the analysis of MI in a multimodal system for 3D object detection.

\subsection{Camera-LiDAR Mutual Information Estimation}
\label{sec:MI_est}
We illustrate the use of InfoMeter by estimating MI between the camera and LiDAR modalities in a 3D object detection system. This investigation provides valuable insights into the role of MI in multimodal learning systems. Specifically, we focus on a state-of-the-art model for this task called FUTR3D (``Fusion Transformer for 3D Detection'')~\cite{futr3d}, which is a end-to-end transformer-based sensor fusion framework for 3D object detection that can be used in autonomous driving scenarios. It provides state-of-the-art detection accuracy on nuScenes~\cite{nuscenes}, a large dataset for autonomous driving. nuScenes consists of about 40,000 data samples. Every sample includes data from three different modalities: 6 cameras, 5 radars, and one LiDAR sensor, accompanied by 3D bounding box annotations.

In FUTR3D, multiple modalities (e.g. camera, LiDAR, and Radar) are fused through a query-based Modality-Agnostic Feature Sampler (MAFS), which consists of feature sampling from each modality followed by modality-agnostic feature fusion~\cite{futr3d}. The integrated data is then utilized by a transformer decoder to perform the 3D object detection task. In our experiments, for the sake of simplicity, we used only the camera and LiDAR modalities. As in~\cite{futr3d}, we employed ResNet-101~\cite{resnet} and VoxelNet~\cite{voxelnet} as the backbones for the camera and LiDAR modalities, respectively. 

\begin{figure*}
\centering
\includegraphics[width=0.8\textwidth]{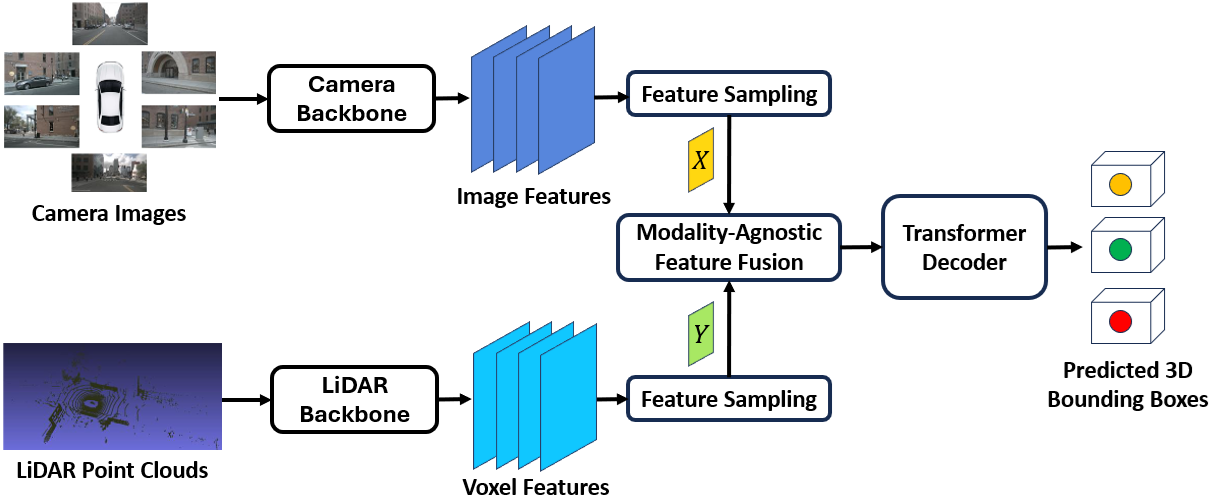}
\caption{The simplified architecture of FUTR3D. In our experiments, $X$ and $Y$ are camera and LiDAR features after feature sampling, which are two single-channel feature maps of the same spatial size. When using 600 queries with an embedding length of 256, these two maps are of size $1\times 600\times 256$.}
\label{fig:futr3d}
\end{figure*}

Fig.~\ref{fig:futr3d} shows the simplified architecture of FUTR3D, and where $X$ and $Y$ are taken from for Camera-LiDAR MI estimation. In our experiments, we used the invertible learned wavelet transform from iWave++~\cite{iwave} as $\phi_X$, $\phi_Y$, and $\phi_{X,Y}$, and the autoregressive entropy model from iWave++ as $h_X$, $h_Y$, and $h_{X,Y}$. Specifically, if $\varphi$ represents the learnable parameters of the invertible transformation and $\theta$ represents the learnable parameters of the entropy estimator, these parameters are learned separately for each branch of the InfoMeter using the entropy estimate as the loss function:
\begin{align}
    (\theta^*_X,\varphi^*_X) &= \argmin_{\theta,\varphi} h_X\left[\phi_X(X;\varphi);\theta\right], \nonumber \\
    (\theta^*_Y,\varphi^*_Y) &= \argmin_{\theta,\varphi} h_Y\left[\phi_Y(Y;\varphi);\theta\right], \label{eq:iWave++_training} \\
    (\theta^*_{X,Y},\varphi^*_{X,Y}) &= \argmin_{\theta,\varphi} h_{X,Y}\left[\phi_{X,Y}(X,Y;\varphi);\theta\right]. \nonumber
\end{align}
Then the entropy estimates are obtained as
\begin{align}
    \widehat{H}(X) &=  h_X\left[\phi_X(X;\varphi^*_X);\theta^*_X\right], \nonumber \\
    \widehat{H}(Y) &=  h_Y\left[\phi_Y(Y;\varphi^*_Y);\theta^*_Y\right], \label{eq:entropy_estimates} \\
    \widehat{H}(X,Y) &=  h_{X,Y}\left[\phi_{X,Y}(X,Y;\varphi^*_{X,Y});\theta^*_{X,Y}\right]. \nonumber
\end{align}

In order to use the iWave++ pipeline to estimate the entropies above, the features need to be adapted appropriately to appear as images.  
In our case, shown in Fig.~\ref{fig:futr3d}, $X$ and $Y$ are feature maps at the output of the feature sampler in MAFS. In the original implementation of FUTR3D, $600$ queries are used and the query embedding length is $256$. Hence, after feature sampling, the resultant feature maps ($X$ and $Y$) are of size $1\times600\times 256$. These maps are rescaled to $[0,255]$ and quantized to the nearest integer to appear as grayscale images. To create the joint representation $(X,Y)$ that appears as an image, feature maps $X$ and $Y$ need to be arranged appropriately. Multiple options have been proposed in the literature for this purpose~\cite{choi_icip}, including tiling and quilting (also known as ``pixel shuffle''). In this work, we use tiling to concatenate the scaled and quantized $X$ and $Y$ to obtain a map of size $1\times600\times 512$ as $(X,Y)$.

\begin{figure*}
\centering
\includegraphics[scale=0.68]{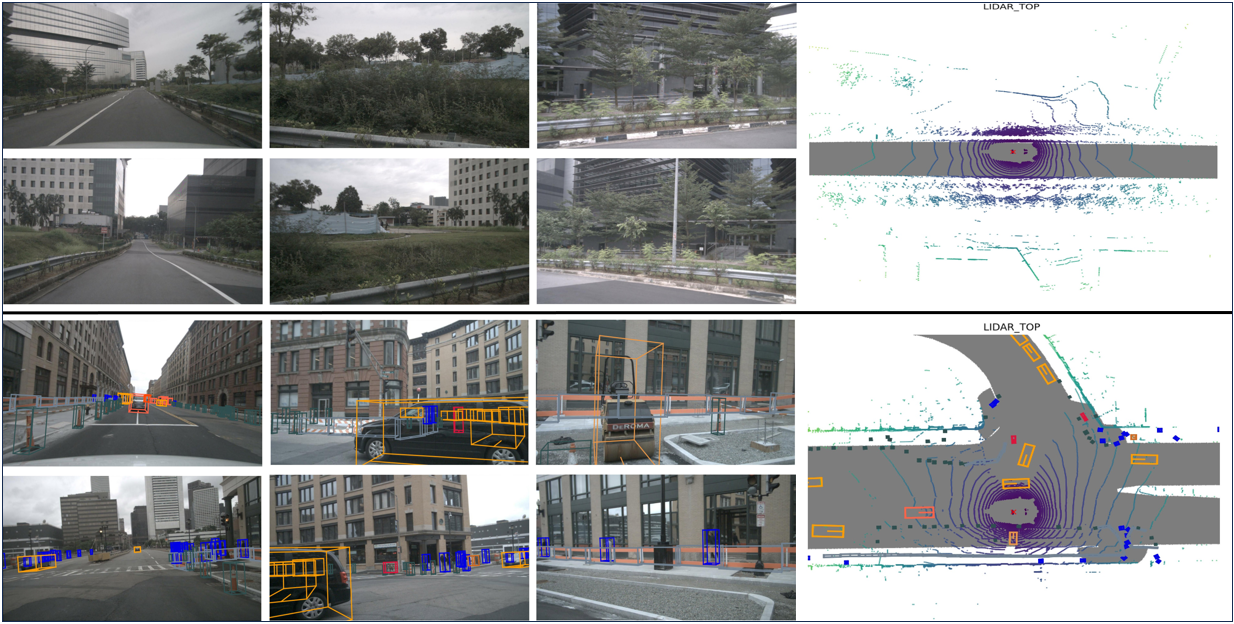}
\caption{Two visual samples from the datasets generated in Scheme 1 showing a low-clutter scene from the first dataset (top) and a high-clutter scene from the second dataset (bottom) along with their ground-truth bounding boxes. Note that the camera modality in nuScenes consists of 6 cameras.}
\label{fig:scheme1}
\end{figure*}

\begin{figure}
\centering
\includegraphics[scale=0.4]{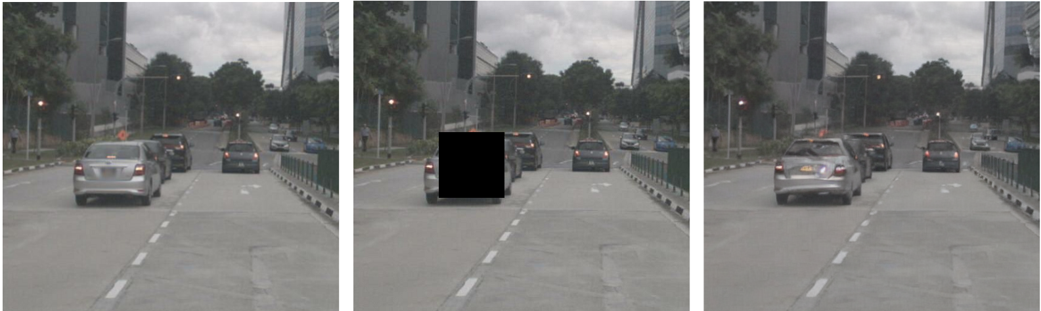}
\caption{A sample image from the dataset generated in Scheme 2. Left: the original image; Middle: the masked image; Right: the inpainted result. As seen from this example, the generative inpainting method used in Scheme 2 was able to effectively reconstruct the masked car.}
\label{fig:scheme2}
\end{figure}

\begin{figure}
\centering
\includegraphics[scale=0.326]{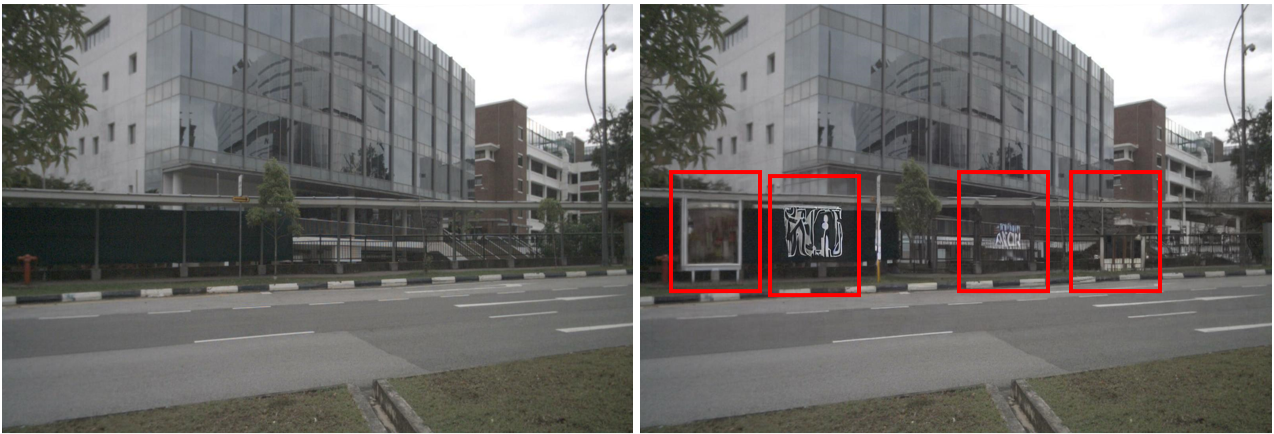}
\caption{An example demonstrating the possible emergence of spurious objects when using a generative inpainting method. Left: the original image; Right: the inpainted result using Scheme 2. The generated spurious objects are highlighted by the red boxes.}
\label{fig:scheme2_dummy}
\end{figure}

\begin{figure*}
\centering
\includegraphics[scale=0.5]{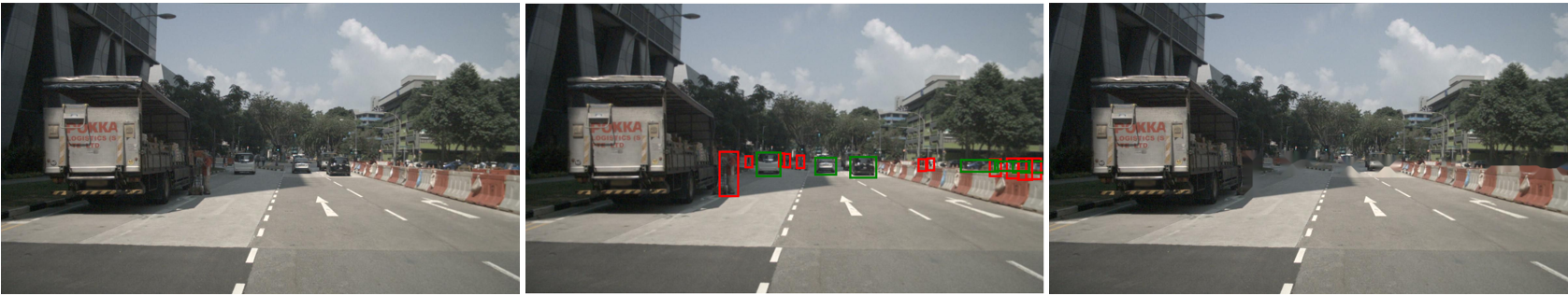}
\caption{A sample image from Scheme 3. Left: the original image; Middle: the objects to be masked; Right: the inpainted result. }
\label{fig:scheme3}
\end{figure*}

\begin{figure*}
\centering
\includegraphics[width=0.8\textwidth]{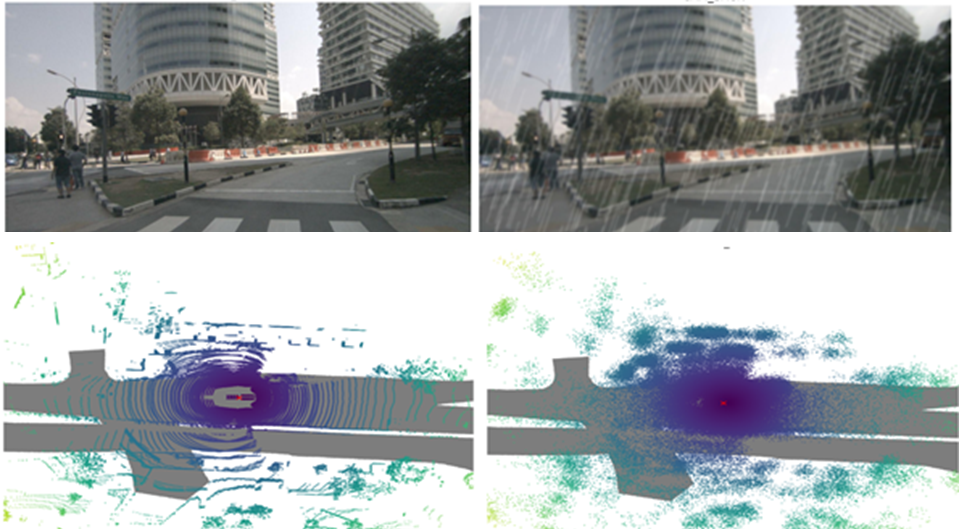}
\caption{A visual example for Scheme 4. The first row shows the original camera image (left) and the distorted ``rainy''image (right). The second row shows the corresponding LiDAR images.}
\label{fig:scheme4}
\end{figure*}

\section{Experiments}
\label{sec:experiments}
In order to study the relationship between MI and model accuracy, we need to create datasets with different amounts of MI between the modalities. We explored four ways of doing this: Schemes~1 through~4 described below. Each scheme involves two training datasets, one with higher MI between camera and LiDAR modalities, and one with lower MI. For each scheme, we first train FUTR3D on the two datasets, obtaining two versions of trained FUTR3D, and then measure their  accuracies on the test set. Then, with FUTR3D backbones frozen, we train InfoMeter as in~(\ref{eq:iWave++_training}) and compute entropy estimates~(\ref{eq:entropy_estimates}) on the respective training sets. From these estimates, the MI estimate is obtained according to~(\ref{eq:MI_estimate}). Then we examine whether higher or lower MI led to better 3D detection accuracy.

\textbf{Scheme 1}: In this scheme, we sorted all samples within the training dataset of nuScenes based on the number of annotations as a simple measure of scene clutter. We then divided the sorted dataset into three subsets of equal size. The subsets with highest and lowest scene clutter were used as two training sets, and the subset with the intermediate scene clutter was used as a test set. The rationale behind this scheme is that in highly-cluttered scenes, occlusion tends to be more pronounced, leading to increased disparity (and therefore lower MI) between the two modalities. The average number of annotations in the low- and high-clutter subsets was approximately 12 and 60, respectively. Fig.~\ref{fig:scheme1} shows a visual sample from each of these two datasets.  

\textbf{Scheme 2}: In this scheme, we generated a new dataset from the training dataset of nuScenes by randomly masking various regions in all six images in the camera modality and then inpainting the masked regions by Stable Diffusion~\cite{stablediff}. Meanwhile, LiDAR data was not modified. The main idea behind this scheme was to deliberately introduce a discrepancy between the two modalities by distorting one of them, resulting in the change in MI. The resulting distorted dataset, in conjunction with the original training dataset of nuScenes, was then used in our experiments. Fig.~\ref{fig:scheme2} shows a sample image along with its masked and inpainted result using this scheme.

\textbf{Scheme 3}: This scheme is similar to the previous one, but employs a different inpainting method. In our experiments, we observed that the generative inpainting method used in Scheme 2 occasionally hallucinates objects (such as new humans or cars) when filling a masked region. An example is shown in Fig.~\ref{fig:scheme2_dummy}. While this is fine for the purpose of changing MI, there is a danger that it introduces too much of a domain shift relative to the test set. Hence, in Scheme 3, we opted for the classic inpainting algorithm from~\cite{classic_inpainting}, which is more conservative and free from object hallucination. Additionally, instead of random region masking, we employed a random object masking strategy - selecting randomly among annotated regions and masking them out. With this scheme, we obtained another distorted dataset, which was utilized alongside the original training dataset of nuScenes in our experiments. Fig.~\ref{fig:scheme3} illustrates a sample image from this dataset. As seen from this example, the inpainted regions are free from hallucinations. Similar to Scheme 2, we kept the LiDAR data intact.

\textbf{Scheme 4}: In this scheme, we created two noisy datatsets by adding ``rain'' noise to both the camera and LiDAR modalities of the nuScenes training set. Rain introduces noise into LiDAR measurements due to multiple reflections from raindrops, which can obscure the 3D contour boundaries of objects in the LiDAR point cloud. To emulate this effect, we used a simple additive Gaussian noise model similar to~\cite{rain_model} to modify the 3D coordinates of the LiDAR point cloud data. For the camera modality, we simulated rainy weather by applying light Gaussian blurring and adding raindrop traces to the camera images. We adjusted noise levels to create approximately equal Signal to Noise Ratio (SNR) in both modalities. This way, we generated a high-noise dataset with SNR of about 21.4 dB in both modalities, and a low-noise dataset with SNR of about 28.1 dB in both modalities. Fig.~\ref{fig:scheme4} shows examples of a noisy image and LiDAR data.

For each scheme, we first train the FUTR3D model on the two datasets.  
For this purpose, we used the training procedure suggested in~\cite{futr3d}. Then we measure the accuracies of the two models in terms of mean Average Precision (mAP)~\cite{futr3d}. After that, we train a separate InfoMeter on each of the two trained models to estimate the MI between the modalities, as shown in~(\ref{eq:MI_estimate})-(\ref{eq:entropy_estimates}).  
Specifically, we trained each iWave++ network using the Adam optimizer~\cite{adam} for 50 epochs. In the first 25 epochs, the learning rate was set to $10^{-4}$, and in the subsequent 25 epochs, it was adjusted to $10^{-5}$. The pre-trained FUTR3D backbones remained frozen during the InfoMeter training.

\section{Results and Analysis}
\label{sec:analysis}
\subsection{Results}
Table~\ref{tab:scheme1} shows the results for Scheme 1. The information quantities (entropy and MI) are in the units of bits per feature element. As seen from these results, the MI on the high-clutter dataset is lower than that in the low-clutter dataset, likely due to the higher number of occlusions that occur in highly cluttered scenes. At the same time, the accuracy of the FUTR3D model trained on the high-clutter dataset is higher. Hence, lower MI leads to higher accuracy in this case. 

\begin{table}[t]
    \centering
    \caption{The MI analysis results for Scheme 1. The units of the estimated entropies and MI are bits per feature element.} 
    \begin{tabular}{c|c|c|c|c|c}
    \toprule
         Dataset & $\widehat{H}(X)$ & $\widehat{H}(Y)$ & $\widehat{H}(X,Y)$ & $\widehat{I}(X;Y)$ & mAP (\%)\\
         \midrule
         Low clutter& 6.0 & 5.8&  6.1& 5.7 & 41.9 \\
         High clutter& 5.6 & 5.6& 6.3 & 4.9 & 55.0 \\
         \bottomrule
    \end{tabular}
    \label{tab:scheme1}
\end{table}

\begin{table}[]
    \centering
    \caption{The MI analysis results for Scheme 2. The units of the estimated entropies and MI are bits per feature element.}
    \begin{tabular}{c|c|c|c|c|c}
    \toprule
         Dataset & $\widehat{H}(X)$ & $\widehat{H}(Y)$ & $\widehat{H}(X,Y)$ & $\widehat{I}(X;Y)$ & mAP (\%) \\
         \midrule
         Original & 5.3 & 5.7 & 5.8 & 5.2 & 64.2  \\
         Distorted & 5.6 &5.7 & 5.9 & 5.4 & 55.6 \\
         \bottomrule
    \end{tabular}
    \label{tab:scheme2}
\end{table}

Table~\ref{tab:scheme2} shows the results for Scheme 2. Note that the entropy of the camera modality ($X$) in the distorted dataset is higher than in the original dataset. This also leads to higher MI estimate but a lower accuracy than when the model is trained on the original dataset. Hence, again, lower MI leads to higer accuracy. 

Table~\ref{tab:scheme3} shows the results for Scheme 3. In this case, the entropy of images is again increased compared to the originals, but not as much as in Scheme 2. The MI estimat in this case is the same as it was with Scheme 2, and again, lower MI leads to higher accuracy.

\begin{table}[t]
    \centering
    \caption{The MI analysis results for Scheme 3. The units of the estimated entropies and MI are bits per feature element.}
    \begin{tabular}{c|c|c|c|c|c}
    \toprule
         Dataset & $\widehat{H}(X)$ & $\widehat{H}(Y)$ & $\widehat{H}(X,Y)$ & $\widehat{I}(X;Y)$ & mAP (\%) \\
         \midrule
         Original & 5.3 & 5.7&  5.8& 5.2 & 64.2 \\
         Distorted& 5.4& 5.7 & 5.7 & 5.4 & 61.3 \\
         \bottomrule
          
    \end{tabular}
    \label{tab:scheme3}
\end{table}

\begin{table}[]
    \centering
    \caption{The MI analysis results for Scheme 4. The units of the estimated entropies and MI are bits per feature element.}
    \begin{tabular}{c|c|c|c|c|c}
    \toprule
         Dataset & $\widehat{H}(X)$ & $\widehat{H}(Y)$ & $\widehat{H}(X,Y)$ & $\widehat{I}(X;Y)$ & mAP (\%) \\
         \midrule
         High noise& 6.5 & 5.7 & 6.7 & 5.5 & 22.8 \\
         Low noise& 6.3 & 5.6& 6.5 & 5.4 & 33.6 \\
         \bottomrule
          
    \end{tabular}
    \label{tab:scheme4}
\end{table}

Finally, Table~\ref{tab:scheme4} shows the results for Scheme 4. We note that the entropy of camera images and the joint entropy of camera and LiDAR modalities have increased compared to earlier schemes, while the entropy estimate of the LiDAR modality has remained approximately the same. This leads to slightly higher MI estimate on the high-noise dataset than on the low-noise dataset. At the same time, the accuracy on the low-noise dataset is higher. Hence, again, lower MI leads to higher accuracy.

\subsection{Analysis}
The above results present four separate lines of evidence that in a multimodal 3D object detection system, lower MI between modalities leads to higher accuracy. This lends support to the ``redundancy'' argument presented in the introduction. Our numerical results are specific to the FUTR3D model trained on camera and LiDAR modalities, but the insight may be useful for other multimodal systems as well. We now offer several interpretations of this insight.

First, recall that MI between modalities quantifies how much information one modality carries about the other. In other words, it is a measure of redundancy between modalities. If this redundancy is high, it means that much of the information carried by a given modality is already included in the other modality, so there is less to be learned from that modality. One way to think of this is to consider that the model being trained may ``overfit'' to the information that is being over-presented to it through different modalities. 

On the other hand, low MI means that the redundancy between modalities is low, so they offer complementary information. By using complementary modalities in training, the model is being exposed to more diverse data, which is known to be beneficial in machine learning~\cite{diversity2019access}. Hence, when describing the roles of different modalities in a multimodal system, it is better to say that they \emph{complement} (rather than \emph{reinforce}) each other. This also suggests that when training multimodal systems, it may be beneficial to incorporate data diversification techniques across modalities, analogous to those used for unimodal training~\cite{diversity2019access}.

\section{Conclusions}
\label{sec:conclusions}
In this paper we introduced InfoMeter, a method for estimating mutual information (MI) built upon state-of-the-art tools from learning-based compression. We then used InfoMeter to estimate MI between camera and LiDAR modalities in a multimodal 3D object detection system. Specifically, we generated multiple training sets with different amounts of MI between camera and LiDAR modalities and trained a recent 3D object detector on them. By comparing the resulting detection accuracy and the MI estimates obtained using the InfoMeter, we found that lower MI leads to higher accuracy. This is in line with the previous findings in unimodal machine learning that data diversity is beneficial, and extends this reasoning to multimodal scenarios. It also suggests that data diversification across modalities during training may lead to improved performance of multimodal systems.

{\small
\bibliographystyle{IEEEtran}
\bibliography{ref}
}

\end{document}